# Stabilization of the epitaxial rhombohedral ferroelectric phase in $ZrO_2$ by surface energy

Ali El Boutaybi, Thomas Maroutian, Ludovic Largeau, Sylvia Matzen, Philippe Lecoeur

*Centre de Nanosciences et de Nanotechnologies, Universite Paris-Saclay, CNRS, Palaiseau, France*

Contact: ali.el-boutaybi@c2n.upsaclay.fr

## Keywords

*ab initio* calculations, epitaxial growth, ferroelectrics, rhombohedral phase, Zirconia

## Abstract

Doped $HfO_2$ and $HfO_2$-$ZrO_2$ compounds are gaining significant interest thanks to their ferroelectric properties in ultrathin films. Here, we show that $ZrO_2$ could be a new playground for doping and strain engineering to increase the thickness in epitaxial thin films. Based on surface energy considerations supported by *ab initio* calculations, we find that pure $ZrO_2$ exhibits a ferroelectric rhombohedral phase (r-phase, with R3m space group) more stable than for the HZO and pure $HfO_2$ cases. In particular, for a thickness up to 37 nm we experimentally evidence a single (111)-oriented r-phase in $ZrO_2$ films deposited on $La_{2/3}Sr_{1/3}MnO_3$-buffered $DyScO_3$(110) substrate. The formation of this r-phase is discussed and compared between $HfO_2$, $ZrO_2$ and HZO, highlighting the role of surface energy.

Since the discovery of ferroelectricity in Si-doped hafnia in 2011 [1], hafnia-based thin films have been widely studied to stabilize the ferroelectric phase, and many polar phases have been suggested or experimentally proved. In polycrystalline thin films, the orthorhombic phase $Pbc2_1$ is usually reported [1][2]. Theoretically, two other polar phases were found for $HfO_2$, an orthorhombic $Pnm2_1$ [3][4] and a rhombohedral R3 [5]. The main problem observed in polar orthorhombic is the necessity of a high number of applied electric field cycles (the so-called wake-up effect) to reach the ferroelectric state [6][7][8]. Moreover, two other issues that are often encountered in these polycrystalline films are the degradation of the ferroelectricity with the increase of film thickness, with a vanishing polarization above a thickness of about 20 nm [1][9][10], and the coexistence of non-polar phases such as monoclinic and tetragonal phases with the polar one [11][12]. Indeed, $HfO_2$ as well as $ZrO_2$ are well known for their structural and chemical similarity[13], and both of them can adopt a wide variety of crystal phases. In bulk form and at room temperature, the stable phase is monoclinic (m-phase, $P2_1/c$). At high temperature, tetragonal phase (t-phase, $P4_2/nmc$) and cubic phase (c-phase, $Fm3m$) are observed for $ZrO_2$ and $HfO_2$ [14][15]. The non-centrosymmetric, ferroelectric orthorhombic $Pbc2_1$ phase (o-phase) can be obtained via transformation of the t-phase under stress [16] [17] [18], or under tensile strain [19]. On the other hand, in 2018, Wei *et al.* [20] demonstrated the rhombohedral symmetry with R3m space group (r-phase) for the first time in epitaxial $Hf_{0.5}Zr_{0.5}O_2$ (HZO) thin films under compressive strain on a $La_{2/3}Sr_{1/3}MnO_3$ (LSMO)-buffered $SrTiO_3$(001) substrate (STO), and no wake-up effect was observed. Strain engineering is thus investigated to stabilize this r-phase on different substrates, such as recently demonstrated by Zheng *et al.* for HZO epitaxial films on ZnO [21].

The influence of surface energy on phase stability has been discussed in polycrystalline films through studies on grain size. It was shown that optimizing the latter could stabilize the orthorhombic $Pbc2_1$ ferroelectric phase in pure $HfO_2$ [22] and $ZrO_2$ [23], and that inserting different interlayers allowed the ferroelectricity to be maintained above 40 nm thickness in HZO [24][25]. Concerning bulk ceramics and polycrystalline films, it is known that for a grain diameter less than 30 nm, the $ZrO_2$ adopts the t-phase [26] [27], whereas the $HfO_2$ has a similar size effect at around 5 nm [11][28]. From this perspective, $ZrO_2$-based epitaxial thin films are promising to increase both film thickness and grain size, which are usually critical for energy storage [29] and optical applications [30][31][32]. Mastering the surface energy balance during growth is thus identified as a critical issue to control the ferroelectric phases of hafnia and zirconia compounds [33][34]. However, very few data are available regarding actual surface energies of the polar and non-polar phases [35], and as far as we know no data concerning surface energy were reported for the rhombohedral polar phase.

The rhombohedral phase in $ZrO_2$ or Partially Stabilized Zirconia (PSZ) has been known for more than two decades. Hasegawa. [36] reported a rhombohedral phase in the abraded surfaces of PSZ and fully stabilized zirconia (FSZ) powders; this phase was only observed in the surface layer under some stress, which can be introduced by polishing and grinding. It was therefore postulated that the rhombohedral phase could only exist in the presence of stress. Also, this rhombohedral phase has been reported as an intermediate phase during cubic and tetragonal–monoclinic transformation [37][38][39]. Nevertheless, in all these earlier studies, no electric properties were reported. R3m and R3 rhombohedral phases were evidenced in HZO epitaxial thin films [20][21][40], and recently, Silva *et al.* [41] reported the same r-phase in a 8 nm-thick $ZrO_2$ thin film deposited on Nb-STO(111). While this polar r-phase is promising for numerous applications based on ultrathin films [42][43][44][45], the issue of its poor thickness-dependent stability is still present when increasing the thickness above 10 nm. In this picture, we will show that rhombohedral $ZrO_2$ films are more stable at high thickness than $HfO_2$ and HZO ones, and thus that $ZrO_2$-rich compounds provide a novel playground for strain engineering and doping schemes towards outstanding ferroelectric properties.

In this letter, we first compare with the support of *ab initio* calculations the stability of the (111)-oriented rhombohedral R3m phase between pure $ZrO_2$, HZO, and pure $HfO_2$. We then study epitaxial $ZrO_2$, HZO, and $HfO_2$ thin films grown by pulsed laser deposition (PLD) on both (110)-oriented $DyScO_3$ (DSO) and (001)-oriented STO substrates, with LSMO buffer. Using X-ray diffraction (XRD), we evidence the presence of the r-phase in $ZrO_2$ thin films, up to about 40 nm thickness, with a clear ferroelectric behavior without wake-up effect.

We performed density functional theory (DFT) calculations to investigate the (111)-oriented rhombohedral stability by employing Quantum Espresso (QE) package [46] with Perdew-Burke-Erzhenhoff (PBE) [47] generalized gradient approximation (GGA) (see supplemental material (SM) for more details [48]). First, we ran our calculations for m-, t-, o-, and r-phases of $ZrO_2$, HZO, and $HfO_2$. The energies of these phases are summarized in Table S1 with a comparison to literature data. Our results agree with previous calculations using different approaches and with the experimental results. For r-$ZrO_2$, we obtained an energy of 140 meV/f.u with respect to the monoclinic energy, which is less than the 154 meV/f.u found for r-$HfO_2$, in good agreement with Ref. [20]. In a second step, the calculations for different surface orientations were performed by constructing 4 layer-thick slabs separated by 15 Å of vacuum (more details in SM. [48]). It is known that the surfaces on each side of a slab may interact through long-range strain fields induced by ionic relaxations [49]. This effect depends somewhat on surface orientation and on the material; for metals, for example, the interlayer relaxations typically fall below the experimental threshold after 3–4 layers [50]. Orlando *et al.* [51] showed the same average value in the case of t-$ZrO_2$. Thus, slabs of 4 layers were used for surface energy calculations (Figure S2) with a 5×5×1 k-point sampling in the surface Brillouin zone (SBZ) [52], and an energy cutoff of 60 Ry.

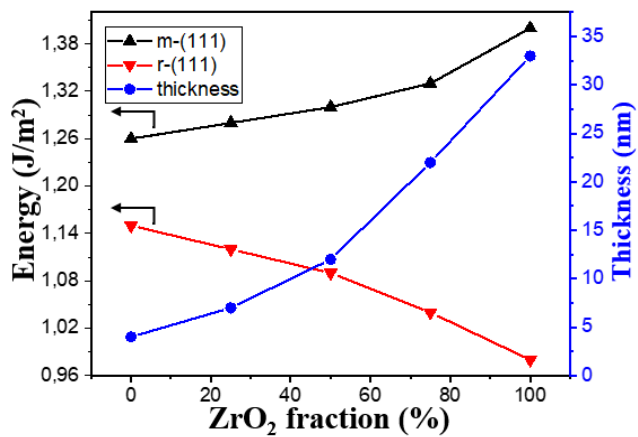

Figure 1. Surface energy of m- and r-phase of $Hf_{1-x}Zr_xO_2$ (x=0, 0.25, 0.50, 0.75, and 1) shown in black and red respectively. The thickness up to which Gibbs energy of the R3m phase is lower than the one of the m-phase is shown in blue.

The surface energy in Joule per square meter is given in Table S2 and Figure 1 for (111)-oriented m- and r-phases at different ratios of $ZrO_2$ in $Hf_{1-x}Zr_xO_2$ composition. The r-(111) showed the lowest surface energy compared to all other calculated surfaces of $HfO_2$, $ZrO_2$, and HZO. As indicated in Figure 1, (111) m-phase increases in energy with increasing $ZrO_2$ content, while the r-(111) phase decreases. This, in turn, increases the energy gap between the m-(111) and r-(111) surfaces, with this gap achieving his maximum in pure $ZrO_2$.

In the literature, the role of the surface energy was not systematically explored in $HfO_2$-$ZrO_2$ epitaxial thin films, especially regarding the stabilization of the rhombohedral phase. In order to assess the relative stability of the m- and r-phases, taking into account the computed surface energies, we use a simple thermodynamic argument based on the Gibbs free energy [33], that was successfully applied to explain the stability of the orthorhombic polar phase in atomic layer deposited (ALD) thin films [11] [33]. Figure 1 shows (in blue) the thickness up to which the Gibbs free energy of the r-phase is lower than the one of the m-phase as a function of $ZrO_2$ content in $Hf_{1-x}Zr_xO_2$. This estimated maximum thickness for the r-phase was calculated with an in-plane compressive strain of 1%, which is close to the estimated strain from our experimental results as shown below. The stability window of r-$ZrO_2$ is up to a thickness of 33 nm, while it is less than 4 nm in pure $HfO_2$. For HZO, rhombohedral phase is stable up to a thickness of about 12 nm, in good agreement with previous results that reported a 10 nm maximum thickness in epitaxial HZO thin films [20]. The (111)-oriented $HfO_2$ R3m phase was discussed elsewhere [53] and found to be stable at a very low thickness (2 layers) compared to the (111)-orthorhombic and (111)-monoclinic. Thus, from our calculations, pure $ZrO_2$ epitaxial thin films are expected to be the most stable regarding the (111)-oriented R3m phase. This makes $ZrO_2$-based thin films very promising to stabilize the rhombohedral polar phase at significantly higher thickness than reported for HZO.

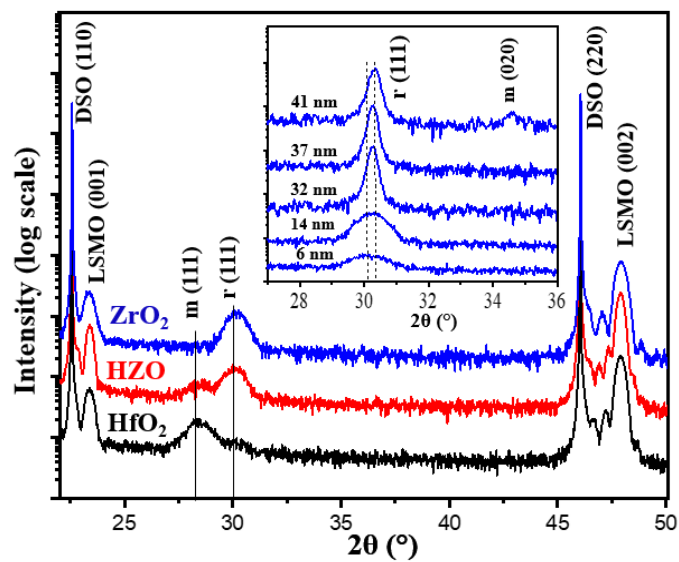

Figure 2. Out-of-plane XRD θ–2θ scans of HZO, $HfO_2$, and $ZrO_2$ films on LSMO-buffered 110-oriented DSO with thicknesses of 14 nm. Insert Figure shows θ–2θ of $ZrO_2$ at different thicknesses, and dashed lines show the 2θ shift of the (111)-$ZrO_2$ diffraction peak.

In order to test our findings experimentally, $ZrO_2$, $HfO_2$, and $Hf_{0.5}Zr_{0.5}O_2$ films were grown by PLD on LSMO buffered DSO and STO substrates. The ferroelectric nature of $ZrO_2$ and HZO was characterized through Positive Up Negative Down (PUND) measurements on capacitor devices with Pt top and LSMO bottom electrodes on DSO substrate (more details are given in SM. [48]). In 14 nm-thick films, a remanent polarizations (2Pr) of 40 μC/cm$^2$ and 41.5 μC/cm$^2$ were measured for HZO and $ZrO_2$, respectively (Figure S4). These results are comparable to the previously reported values in epitaxial HZO with the same range of thickness [20][54] [55]. At 37 nm thickness, a 2Pr of 22 μC/cm$^2$ is measured for pure $ZrO_2$ (Figure S4c), hinting at a persistent ferroelectricity.

Structural analyses were performed by XRD with Panalytical X'pert Pro diffractometer for out-of-plane θ–2θ scans (Figure 2) and Rigaku SmartLab diffractometer equipped with a rotating anode for in-plane measurements and pole figure (Figure 3). Figure 2 gives θ–2θ patterns of HZO, HfO$_2$, and ZrO$_2$ films at a thickness of 14 nm. The highest peaks correspond to the (110)-oriented orthorhombic DSO substrate. At the right and close to the substrate peaks, are the LSMO peaks corresponding to (001) pseudo-cubic orientation. The thickness of LSMO is about 25 nm, measured by X-ray reflectivity. For 2θ between 30.09°-30.27°, is found the (111) peak of the HfO$_2$, HZO, and ZrO$_2$ films. Note that, the diffracted peak around 30° is referred to the (111) orthorhombic Pbc2$_1$ in polycrystalline thin films [1], and even in some epitaxial thin films [54]. In the case of polycrystalline films this peak is usually found at a slightly higher 2θ value (around 30.5°) compared to epitaxial thin films [10]. Here, the (111) diffracted peak will be ascribed to the rhombohedral phase, as demonstrated below through complimentary XRD analyses. In the case of pure HfO$_2$, a peak at 2θ around 28.3° is observed with higher intensity than r-(111); this peak corresponds to (111) m-phase [56]. This m-(111) peak decreases in intensity in HZO film and disappears in pure ZrO$_2$ film. The inset of Figure 2 shows the XRD data of pure ZrO$_2$ films at different thicknesses. The r-phase in the film appears compressively strained, as the (111) peak is shifted towards smaller angles for the thinner films, indicating that the out-of-plane parameter is expanded compared to the relaxed one at 41 nm. From this thickness-dependent shift, a compressive strain of around 1% was extracted [48]. Additionally, a diffraction peak at 34.6° attributed to m-(020) plane is observed in the 41 nm-thick ZrO$_2$ film [56], while no monoclinic phase is observed for thicknesses up to 37 nm. Noteworthy, an increase in compressive strain higher than 1% could increase the thickness of the pure r-ZrO$_2$ film much higher than 37 nm.

In line with our theoretical considerations, both our experimental composition and thickness series confirm that a polar (111)-oriented phase remains stable at high thickness upon increasing the ZrO$_2$ content in HfO$_2$-ZrO$_2$ compounds, allowing to reach a thickness close to 40 nm in pure ZrO$_2$ on DSO substrate. We now present the additional XRD measurements that unambiguously identify the rhombohedral phase in our ZrO$_2$ thin films, as previously reported for HZO films [20]. Note that due to the very low intensity of the r-phase related peak in our pure HfO$_2$ films (Figure 2, black), we could not perform a similar analysis for them.

Figure 3a shows a pole figure for a 14 nm-thick ZrO$_2$ film on DSO substrate measured at 2θ= 30.19°; 12 radial peaks were found, corresponding to four variants of ZrO$_2$. For one r-ZrO$_2$ variant, only three peaks are expected at a radial angle χ = 71° [48]. The three inclined planes diffract at a different angle than the (111) plane (surface plane of the thin film); which gives a multiplicity of 3:1, that is a feature of rhombohedral symmetry. Figure 3b shows θ-2θ scans for {-111} planes family of one rhombohedral variant (selected in red in pole figure) inclined relatively to the sample surface: A clear shift between these peaks collected at χ ~ 71° and the one corresponding to the (111) plane parallel to the surface is observed, confirming the rhombohedral symmetry [20]. The shift between the central peak and the inclined ones is around Δ(2θ) = 0.31°, and the same shift was observed for the other variants (Figure S5). Importantly, with d$_{hkl}$ the interplanar spacing of (hkl) planes, the d$_{111}$– d$_{11-1}$ difference depends on the α angle of the rhombohedral phase. An angle α = 89.40° was reported in HZO thin film with a thickness of 5.9 nm [40], and α = 89.5° in ZrO$_2$-3% Y$_2$O$_3$ powder [38], which give a Δ(2θ) between the r-(111) and r-{-111} planes of about 0.42° and 0.36°, respectively. According to the shift observed in Figure 3b, the estimated angle for 14 nm-thick pure ZrO$_2$ r-phase is α = 89.56° with lattice parameters a = b = c ≈ 5.089Å. The values of α and of the corresponding out-of-plane d$_{111}$ at different thicknesses of ZrO$_2$ films are reported in Figure 3c. They exhibit a clear dependence on thickness, characteristic of the relaxation of an in-plane compressively strained r-phase: The angle α increases toward 90° and $d_{111}$ decreases as the film thickness increases. This relaxation is consistent with the decrease of polarization measured in $ZrO_2$ films between 14 nm and 37 nm thickness (Figure S4). Indeed, compressive strain was shown to have a strong impact on the ferroelectric properties of the r-phase [20][40][53].

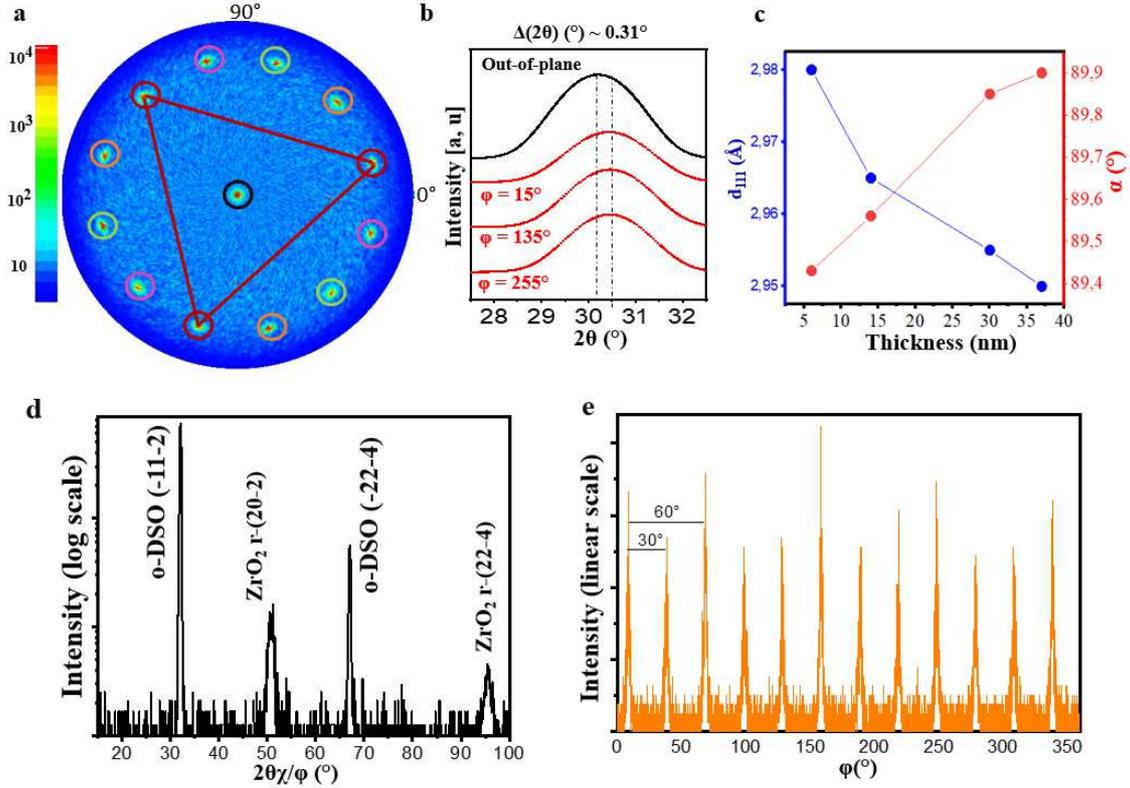

Figure 3. a) Pole figure around the (111) peak at 2θ = 30.19° of a 14 nm-thick $ZrO_2$ film. The radial angle χ varies between 0° and 90°, the azimuthal angle φ is in the 0°–360° range. b) θ-2θ scans of one r-variant (red circles in pole figure) revealing a Δ(2θ) shift of 0.31° between diffraction peaks corresponding to out-of-plane and 71°-inclined {111} planes. c) out-of-plane $d_{111}$ and rhombohedral angle α at various thicknesses d) In-plane 2θχ/φ scan along <-11-2> DSO in-plane azimuthal direction. e) In-plane φ –scan around 2θχ=50.54°.

In order to get further insights into the epitaxial relationships of $ZrO_2$ films on LSMO-buffered DSO and STO, and to be more sensitive to the eventual presence of minority phases, we performed a series of in-plane XRD measurements. These in-plane measurements were done only for HZO (SM. [48]) and pure $ZrO_2$ films; as in pure $HfO_2$, the dominant phase is monoclinic. First, we searched the (-11-2) plane of orthorhombic DSO (o-DSO) substrate that diffracts at around 2θ = 32.09° (called 2θχ in in-plane geometry) by rotating the sample in-plane along the φ angle [57]. That allows to fix the offset between 2θχ and φ. Then, in-plane 2θχ/φ scans were performed along the azimuthal substrate directions, as given in Figure 3d. In addition to the substrate peaks, two other peaks are present at 2θχ = 50.54° and 2θχ = 95.57°, which correspond to the rhombohedral (20-2) and (22-4) planes, respectively. As the (22-4) plane is rotated by 30° relative to the (20-2) plane; in fact, from Figure 3d, two rhombohedral variants rotated by 90° against each other are observed (Figure S7c). In order to determine the in-plane symmetry of the $ZrO_2$ phase, we performed in-plane φ scans with the detector fixed at 2θχ = 50.54°. The φ–scan is shown in Figure 3e, revealing 12 peaks with a separation Δφ = 30°, confirming the six-fold symmetry (hexagonal notation) of $ZrO_2$ thin film. The φ–scan for 2θχ = 95.57° and for the substrate are given in SM (Figure S6), also indicating a six-fold symmetry of the peak around 95.57°. Similar results were obtained for HZO thin film (Figure S7 in SM. [48]). Finally, we note that no LSMO peak was detected in the in-plane scans, neither on DSO nor on STO substrate, in agreement with a fully strained LSMO buffer layer [58].

Furthermore, growth of $ZrO_2$ and HZO films on STO substrate were also tested and the same structural analyses as discussed for DSO substrate were performed. Interestingly, a tetragonal phase has been detected along the r-phase by in-plane measurements (Figure S9 and S10). The in-plane lattice parameters calculated from XRD for this phase are a = b ≈ 5.088 Å, close to the value reported for the relaxed t-$ZrO_2$ [59][60]. This t-phase was not observed on DSO substrate neither in out-of-plane nor in in-plane XRD scans, even at the highest studied thicknesses of $ZrO_2$ thin films (Figure 2), hinting at an effect of the different strain caused by DSO compared to STO substrate on t-phase stability.

In conclusion, the rhombohedral phase in $HfO_2$-$ZrO_2$ compounds was systematically studied using first-principles calculations and experiments. We evidenced the stability of (111) oriented r-phase in $ZrO_2$ up to the 30-40 nm thickness range from surface energy considerations, and demonstrated experimentally the stabilization of this pure rhombohedral phase in a 37 nm-thick $ZrO_2$ thin film deposited on LSMO buffered DSO (110) substrate, with a remanent polarization $2P_r$ of 22 µC/cm$^2$. On STO (001) substrate, a relaxed tetragonal phase was found to coexist with rhombohedral phase in HZO and pure $ZrO_2$ films. We believe that $ZrO_2$-rich epitaxial thin films open new routes in terms of strain engineering and doping for optimized ferroelectric properties and thicknesses higher than the ones reported in this work.

Acknowledgements: This work has received support from the French national research agency (ANR) under project FOIST (N°ANR-18-CE24-0030), and from the French national network RENATECH for nanofabrication. The authors thank Michel Rerat (IPREM-Pau University, France) for helpful discussions on our *ab initio* calculations.

# Supplemental material

# Stabilization of the epitaxial rhombohedral ferroelectric phase in ZrO$_2$ by surface energy

Ali El Boutaybi, Thomas Maroutian, Ludovic Largeau, Sylvia Matzen, Philippe Lecoeur

*Centre de Nanosciences et de Nanotechnologies, Université Paris-Saclay, CNRS, Palaiseau, France*

Contact: ali.el-boutaybi@c2n.upsaclay.fr

**DFT calculations**

We used the Quantum Espresso (QE) package for our first-principles calculations with a 12-atom cell for HfO$_2$ and ZrO$_2$ structures. We tested our calculation results by reproducing some of the previous results (Ref[1]), the calculations have been performed for orthorhombic Pbc2$_1$ (o-phase), monoclinic P2$_1$/c (m-phase), tetragonal P4$_2$/nmc (t-phase), and rhombohedral R3m (r-phase) (Figure S1). For R3m phase, we considered a 12-atom cell for HfO$_2$ with 3-fold-symmetry and we ran an optimization to find the positions of the 12 atoms corresponding to the minimum energy while keeping 3-fold symmetry. Then, we replaced some of the Hf atoms with Zr atoms, according to the targeted composition Hf$_{1-x}$Zr$_x$O$_2$ (x=0, 0.25, 0.5, 0.75, and 1). A cell optimization (lattice parameters and volume) of R3m phase was allowed for all compositions, with a fixed angle at 89.5° as given in Refs [2] [3].

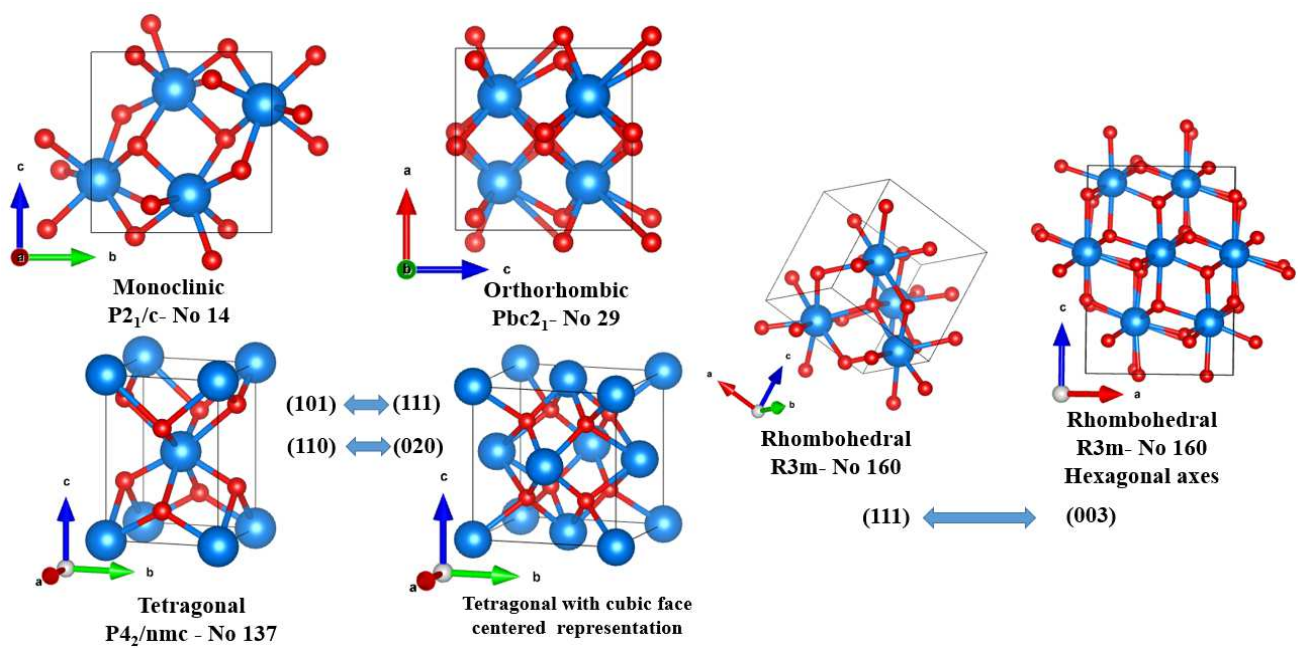

Figure S1. Different phases of ZrO2. The tetragonal cubic face-centered is used in DFT calculation to facilitate the data comparison and for 111-oriented surface energy calculation (Table S2).

Table S1. Comparison of lattice constants a, b, and c in Å and total energy difference ΔU** in meV/f.u. for $HfO_2$, $ZrO_2$ and HZO.

| | Energy (meV) | Volume (Å) | a(Å) | b(Å) | c(Å) | Technique and Ref |
|---|---|---|---|---|---|---|
| m-$HfO_2$ | / | 138 | 5.11 | 5.17 | 5.29 | Exp[4] |
| | / | 138.2 | 5.11 | 5.19 | 5.28 | PBE* |
| | / | 137.1 | 5.12 | 5.20 | 5.28 | LDA[1] |
| | / | - | 5.13 | 5.19 | 5.30 | GGA[5] |
| t-$HfO_2$ | / | … | 5.06 | 5.06 | 5.20 | Exp[6] |
| | 122 | 138.8 | 5.05 | 5.05 | 5.20 | PBE* |
| | 92 | 129.5 | 5.03 | 5.03 | 5.12 | LDA[1] |
| o-$HfO_2$ | / | … | … | … | … | Exp |
| | 90 | 132.2 | 5.01 | 5.045 | 5.23 | PBE* |
| | 83.4 | - | 5.06 | 5.09 | 5.27 | PBE[7] |
| | 83 | - | - | - | - | PBE[8] |
| | 62 | 132.1 | 5.02 | 5.04 | 5.22 | LDA[1] |
| r-$HfO_2$ | 154 | 132.3 | 5.097 | 5.097 | 5.097 | PBE* |
| | 158 | - | 5.09 | 5.09 | 5.09 | PBEsol[9] |
| m-$ZrO_2$ | / | 140.3 | 5.15 | 5.20 | 5.32 | Exp[10] |
| | / | 140.6 | 5.16 | 5.20 | 5.33 | PBE* |
| | / | 138.2 | 5.11 | 5.20 | 5.28 | LDA[1] |
| | / | - | 5.17 | 5.23 | 5.34 | LDA[11] |
| t-$ZrO_2$ | / | … | 5.11 | 5.11 | 5.27 | Exp[12] |
| | 90 | 138.8 | 5.10 | 5.10 | 5.26 | PBE* |
| o-$ZrO_2$ | / | … | 5.07 | 5.08 | 5.26 | Exp[10] |
| | 61 | 138.8 | 5.09 | 5.12 | 5.32 | PBE* |
| r-$ZrO_2$ | / | - | 5.089 | 5.089 | 5.089 | Exp†* |
| | 140 | 135.5 | 5.13 | 5.13 | 5.13 | PBE* |
| m-HZO | / | 139 | 5.14 | 5.19 | 5.32 | PBE* |
| | / | 137.6 | 5.11 | 5.18 | 5.28 | LDA[1] |
| o-HZO | | | 5.01 | 5.05 | 5.24 | Exp[13] |
| | 68 | 135.2 | 5.05 | 5.08 | 5.27 | PBE* |
| | 59 | 132.8 | 5.03 | 5.05 | 5.23 | LDA[1] |
| r-HZO | 152 | - | 5.09 | 5.09 | 5.09 | Exp* |
| | 152 | 133.5 | 5.11 | 5.11 | 5.11 | PBE* |
| | 154 | - | - | - | - | PBEsol[9] |

*This work, ** ΔU is given with respect to the monoclinic phase (ground state)

† These parameters are calculated from a 14-nm thick $ZrO_2$ thin film.

Once these calculations were done for all the considered phases (Table S1), slabs of 4 layers and 48 atoms were created at fixed lattice parameters with different orientations, with a focus on (111)-orientation as it is the most present in our epitaxial thin films. The slab geometries of (111)-oriented rhombohedral and monoclinic phases are given in Figure S1. The slabs are with nonpolar surfaces for all investigated phases, comprised of an integer number of $ZrO_2$ formula units.

The surface energy was computed following the relation (1) below, and the results are summarized in Table S2.

$$\gamma_{slab}^{\alpha}(t) = \frac{1}{2 A_{slab}^{\alpha}} [E_{slab}^{\alpha}(t) - n E_{bulk}^{\alpha}] \quad (1)$$

Where $E_{slab}^{\alpha}$ is the DFT energy of the slab with thickness t, $A_{slab}^{\alpha}$ is the slab area, $n$ is the total number of formula units, and $E_{bulk}^{\alpha}$ is the bulk energy. The superscript $\alpha$ denotes the phase.

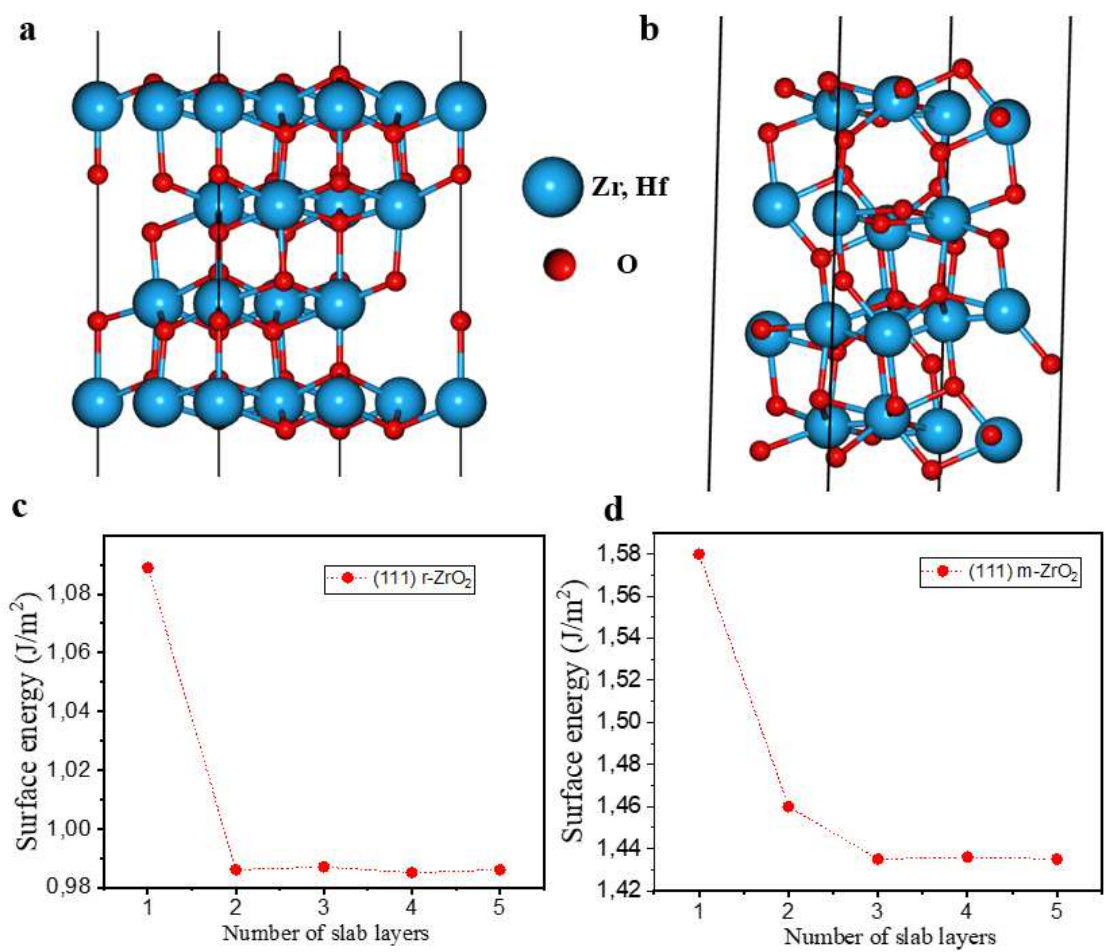

Figure S2. Periodically repeated slab geometry with 48 atoms and convergence test with the number of slab layers for surface energy calculations of r-phase (a and c) and m-phase (b and d), respectively.

Table S2. Surface energy for $HfO_2$, $ZrO_2$, and HZO in $J/m^2$.

|  | This work | Literature |
|---|---|---|
| m-$HfO_2$ | | |
| 001 | 1.82 | 1.55[14] |
| 111 | 1.26 | 1.25[14] |
| t-$HfO_2$ | | |
| 001 | 1.26 | 1.21[14] |
| 111 | 1.25 | 1.12[14] |
| o-$HfO_2$ | | |
| 111 | 1.6 | - |
| **r-$HfO_2$** | | |
| **R3m 111** | **1.15** | **-** |
| m-$ZrO_2$ | | |
| 001 | 1.80 | 1.802[15] |
| 111 | 1.42 | 1.246[15] |
| t-$ZrO_2$ | | |
| 001 | 1.19 | 1.57[15] |
| 111 | 1.21 | 1.24[15] Exp 1.23[16] |
| o-$ZrO_2$ | | |
| 111 | 1.49 | - |
| **r-$ZrO_2$** | | |
| **R3m 111** | **0.98** | **-** |
| m-HZO | | |
| 001 | 1.87 | - |
| 111 | 1.31 | - |
| t-HZO | | |
| 001 | 1.25 | - |
| 111 | 1.24 | - |
| o-HZO | | |
| 111 | 1.54 | - |
| **r-HZO** | **-** | **-** |
| **R3m 111** | **1.09** | **-** |

**Thickness-dependent Gibbs free energy**

The calculated surface energies (Table S2) are integrated in a simple thermodynamic model (Ref [1]) to obtain a thickness-dependent Gibbs free energy for the (111)-oriented monoclinic, rhombohedral and orthorhombic phases. Note that the (111)-oriented o-phase is more stable than the m-phase below 20 nm thickness, but its Gibbs free energy is higher than the one of the r-phase (Figure S3, blue).

Gibbs free energy is calculated as follows [1][17]:

$$\mathbf{G} = \mathbf{F} + \gamma \mathbf{A} + \mathbf{V}\,\sigma_{ii}\varepsilon_{ii} \qquad i = x; y; z \qquad (2)$$

Where $\gamma$, A, V, $\sigma_i$, and $\varepsilon_i$ are surface energy, surface area, volume, stress tensor, and strain tensor, respectively. The Helmholtz free energy F for the monoclinic phase is taken as zero at 298.15 K. Then, multiplying the equation (2) by (1/V) will give G/V as a function of ($1/t_{film}$), where $t_{film}$ is the film thickness (Figure S3). The compressive strain was estimated from the measured relaxation of the $ZrO_2$ films with increasing film thickness using the equations (3) and (4) [18].

$$\varepsilon_{oop} = \frac{-2\nu}{1-\nu} \varepsilon_{in} \quad (3)$$

Where $\varepsilon_{oop}, \varepsilon_{in}$, and $\nu$ are out-of-plane strain, in-plane strain, and Poisson coefficient, respectively. For a Poisson coefficient taken as 0.30 [19], then $\varepsilon_{oop} \approx -\varepsilon_{in}$. In-plane strain is thus estimated from $\varepsilon_{oop}$, calculated using the following equation:

$$\varepsilon_{oop} = \frac{c_f - c_{rf}}{c_{rf}} \quad (4)$$

Where $c_f$ is the out-of-plane parameter of strained r-phase at 6 nm thickness ($c_f$ = 2.98 Å), and $c_{rf}$ is the out-of-plane parameter of the relaxed r-phase at 41 nm thickness ($c_{rf}$ = 2.95 Å). Together with equation (3), this gives an estimated in-plane compressive strain of 1%. We used the out-of-plane parameter of the relaxed phase instead of the bulk phase, as no experimental data are available for bulk rhombohedral pure $ZrO_2$.

Then, the elastic strain energy was calculated using this relation:

$$\sigma_i = \frac{E}{(1-2\nu)(1+\nu)} [(1-\nu)\varepsilon_i + \nu(\varepsilon_j + \varepsilon_k)] \quad (5)$$

Where the elastic modulus $E$ is taken as 180 GPa for all phases and compositions [20], and the i,j,k indexes stand successively for the x,y (in-plane) and z (out-of-plane) directions.

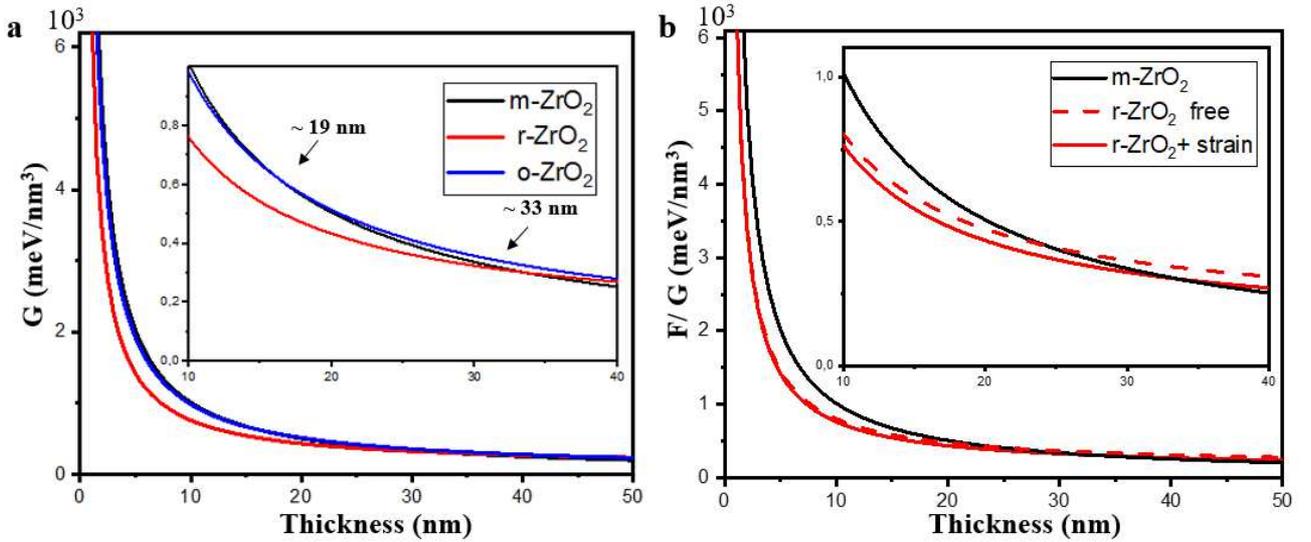

Figure S3. a) Gibbs free energy (with 1% compressive strain, see text) of m- and polar o-, and r-phases as a function of film thickness for an epitaxial $ZrO_2$ film. b) Free energy F + γA (without compressive strain) of r-phase compared to Gibbs energy (with strain) of m- and r-phase as a function of film thickness.

From Figure S3a, we observe that the Gibbs energy of rhombohedral (R3m) is lower than that of monoclinic phase when the thickness is less than 33 nm. Above this value, the monoclinic phase becomes more stable than the (111)-oriented rhombohedral. While the Gibbs energy of orthorhombic (Pbc2$_1$) phase is lower than the one of monoclinic phase when the thickness is less than 19 nm, it is still higher than for the rhombohedral. Note that in this simple Gibbs energy picture, strain-free r-$ZrO_2$ is still the most stable phase up to about 25 nm (Figure S3b). Also, the surface energy difference between HZO, $ZrO_2$ and $HfO_2$ is enough to account for the experimentally observed hierarchy of r-phase stability in these compounds, without invoking grain boundaries or other interface effects.

**Thin-film synthesis**

Thin films of $ZrO_2$, $HfO_2$, and HZO were grown by pulsed laser deposition (PLD) on LSMO-buffered DSO(110) and STO(001) substrates. A laser fluence of 1.2 J/cm$^2$ (wavelength of 248 nm), a repetition rate of 2 Hz, and growth temperature of 760 °C were used for all depositions. The oxygen pressure was 120 mTorr for LSMO and 75 mTorr for $Hf_{1-x}Zr_xO_2$ (x=0, 0.5, and 1) films, respectively. Right after growth, the films were cooled down to room temperature under an oxygen pressure of 300 Torr, at a rate of 10 °C/min.

**P-E characterization**

On top of $Hf_{1-x}Zr_xO_2$/LSMO/DSO, Platinum (Pt) top-electrode pads with square shapes ranging from 10x10 to 200x200 µm$^2$ were defined by photolithography. Two photolithography levels were used, one for SiN the second for Pt deposition (Figure S4d). Sputtering was used for SiN dielectric (Plassys MP800S) and Pt metal (Alliance Concept AC450) depositions. Pt/$ZrO_2$/LSMO and Pt/HZO/LSMO capacitors were characterized using a ferroelectric tester (AiXACCT, TF analyzer 1000). The film ferroelectric response was obtained via Positive Up Negative Down (PUND) measurements, allowing for the switching current to be separated from other contributions (Figure S4).

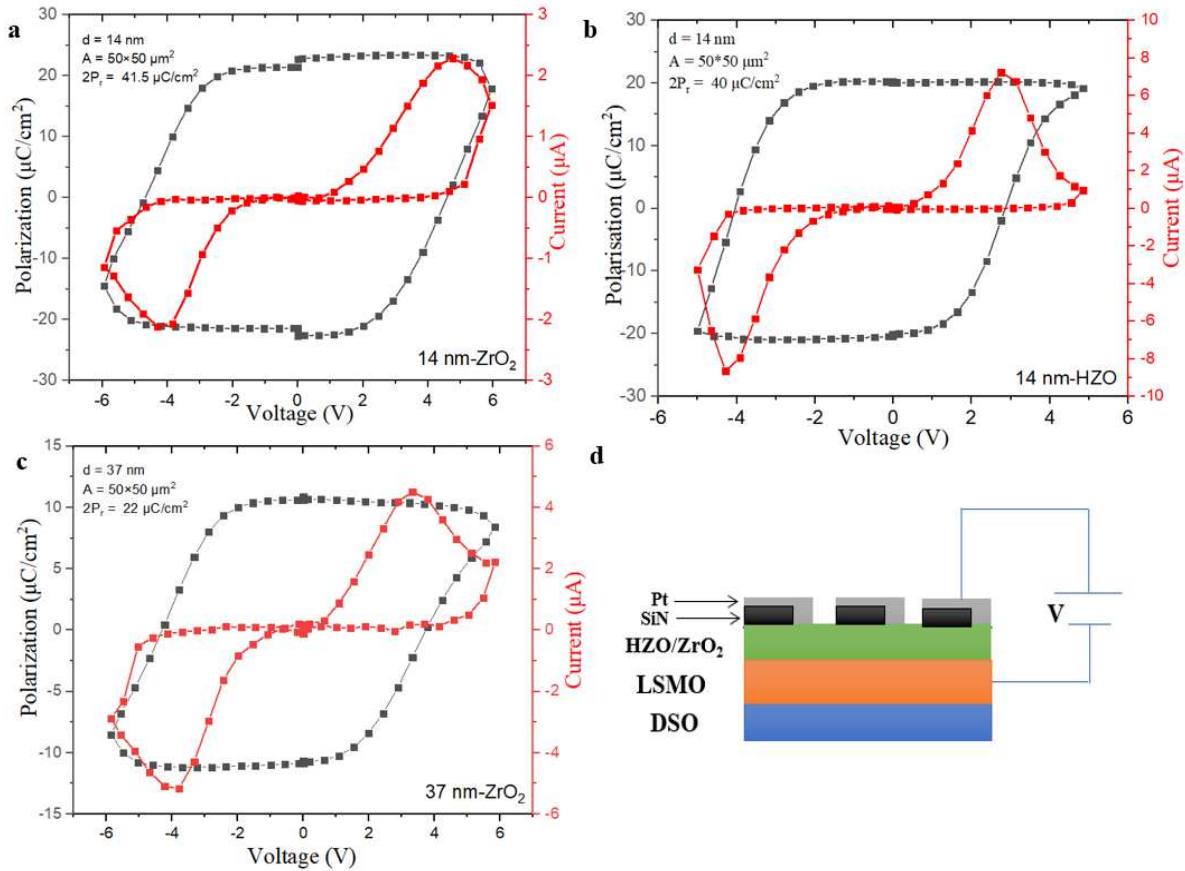

Figure S4. a, b) Polarization–voltage hysteresis loops of 14 nm-thick $ZrO_2$ and HZO on DSO substrate. c) Polarization–voltage hysteresis curves of 37 nm-thick $ZrO_2$ film on DSO substrate. d) Schematic view of metal/film/metal capacitor configuration for P-E characterization.

## Structural characterization

Structure and epitaxial orientations of the thin films were characterized by X-ray diffraction (XRD). Panalytical X'pert Pro diffractometer was used for θ-2θ characterizations, while pole figures and φ scans (azimuthal angle) were obtained with a Rigaku SmartLab. Both diffractometers were operated in parallel beam configuration with monochromated Cu Kα1 radiation (wavelength of 1.54059Å).

ZrO$_2$ and HZO on DSO substrate

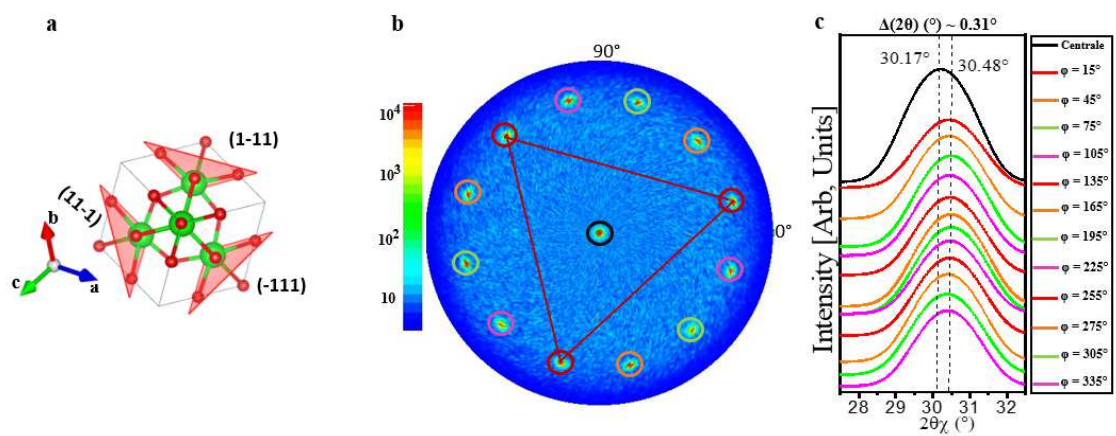

Figure S5. a) Rhombohedral phase with three planes in radial directions at χ around 71° (red color). b) Pole figure around the (111) peak at 2θ = 30.19° of a 14 nm-thick ZrO$_2$ film. The radial angle χ varies between 0° and 90°, the azimuthal angle φ is in the 0°–360° range. c) θ-2θ scans of the 13 peaks observed in the pole figure revealing a Δ(2θ) shift of 0.31° between diffraction peaks corresponding to out-of-plane and 71°-inclined {111} planes.

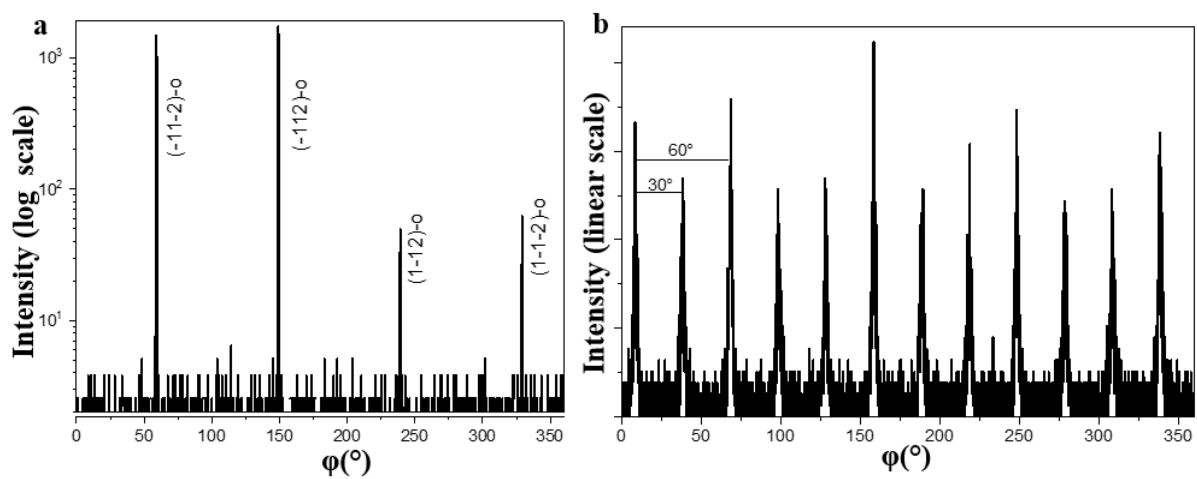

Figure S6. a) In-plane φ-scan of {11-2} DSO substrate collected at 2θχ = 32.09°. b) In-plane φ-scan of ZrO$_2$ film at 2θχ = 95.54° corresponding to the peak observed in Figure 3a.

Figure S6a confirms the pseudo-cubic four-fold symmetry of the DSO substrate. Figure S6b confirms the six-fold symmetry of the peak around 95.54°; from this figure, four rhombohedral R3m variants are observed.

Figure S7 presents in-plane diffraction scans of HZO on DSO substrate. The tetragonal phase was not detected; only the rhombohedral phase is present. Two r-HZO variants are observed in Figure S7a as illustrated in Figure S7c.

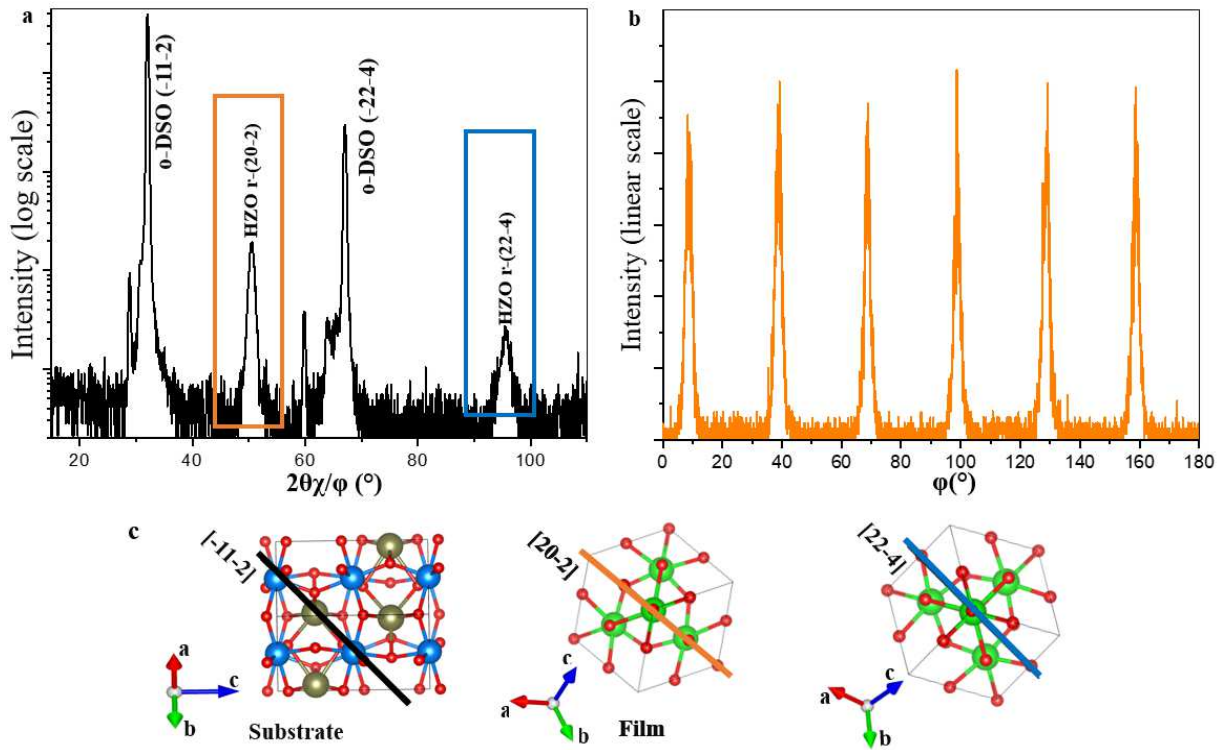

Fig S7. a, b) In-plane 2θχ/φ scan of HZO film of 14 nm thickness on DSO substrate. b) in-plane φ-scan (0-180°) at 2θ = 50.53° showing the hexagonal symmetry. c) Schematic top view of the parallel planes to (-11-2) of the DSO substrate showing two rhombohedral phases rotated by 90° relative to each other.

## ZrO$_2$ and HZO on STO substrate

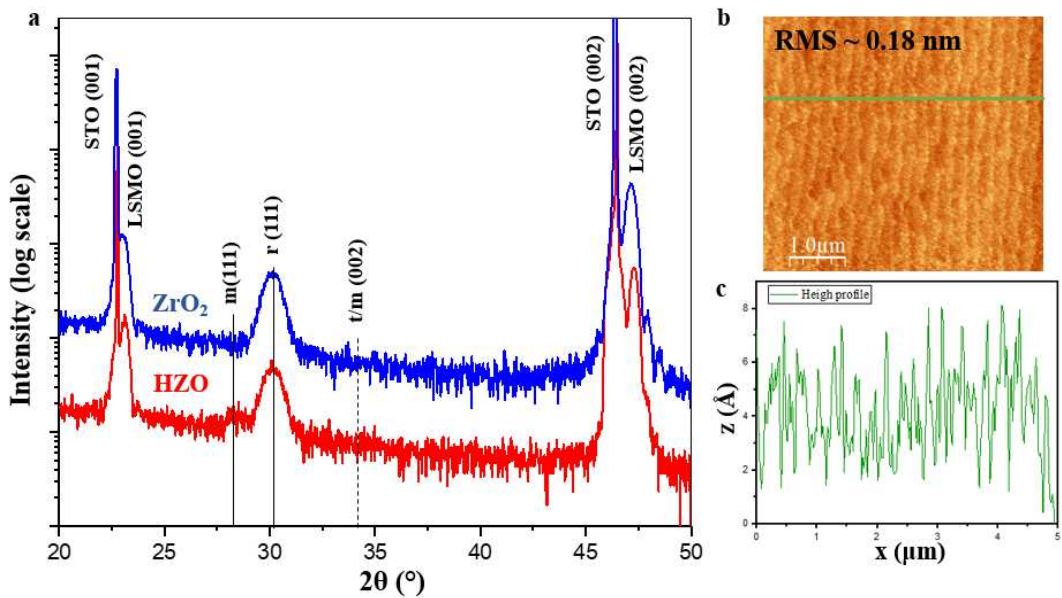

Figure S8. a) XRD structural characterization of 14 nm-thick ZrO$_2$ and HZO films on LSMO-buffered STO (001); b) AFM image of the ZrO$_2$ film with root mean square (RMS) roughness and c) height profile along the horizontal green line.

As shown in Figure S8, at 14 nm thickness a clear monoclinic peak around 28.3° is observed for HZO, while none is seen for ZrO$_2$. The dashed line indicates the angle at which would be the tetragonal peak corresponding to the t-phase detected in in-plane measurements (Figures S9 and S10). The surface topography of ZrO$_2$ was characterized by atomic force microscopy (AFM, Innova from Bruker) in contact mode. Representative 5 x 5 µm² image and extracted height profile [21] are shown in Figure S8.

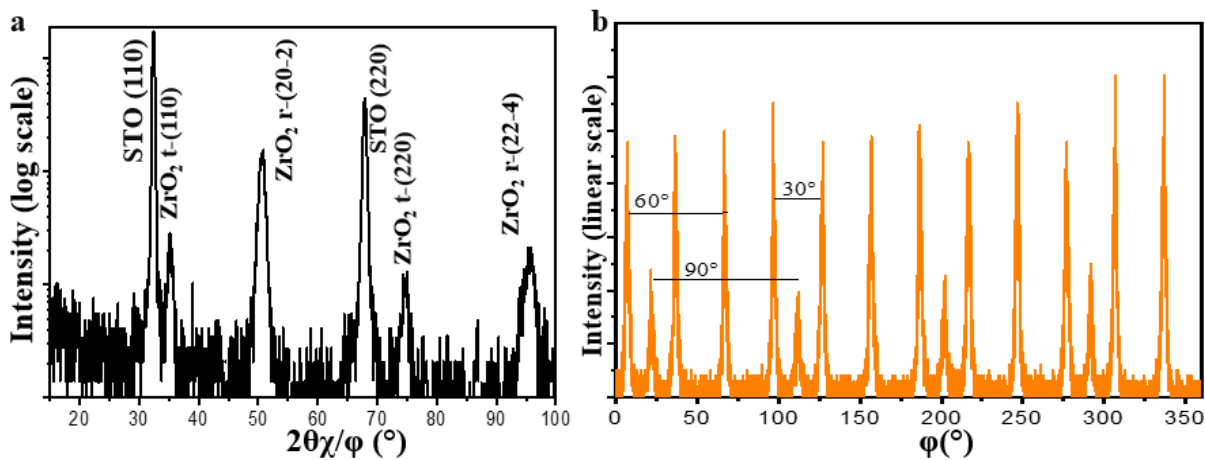

Figure S9. a) In-plane 2θχ/φ scan along <110> of STO in-plane azimuthal direction. b) In-plane φ –scan around 2θχ=50.67°. These measurements have been performed on the film of 14-thick ZrO₂ film (Figure S8).

In the case of ZrO₂, Figure S9b shows 12 peaks corresponding to four r-phase variants, but also reveals the four-fold symmetry of a tetragonal phase in the film. This t-phase ZrO₂ is not detected in the out-of-plane 2θ as illustrated in Figure S8, for the same ZrO₂ film.

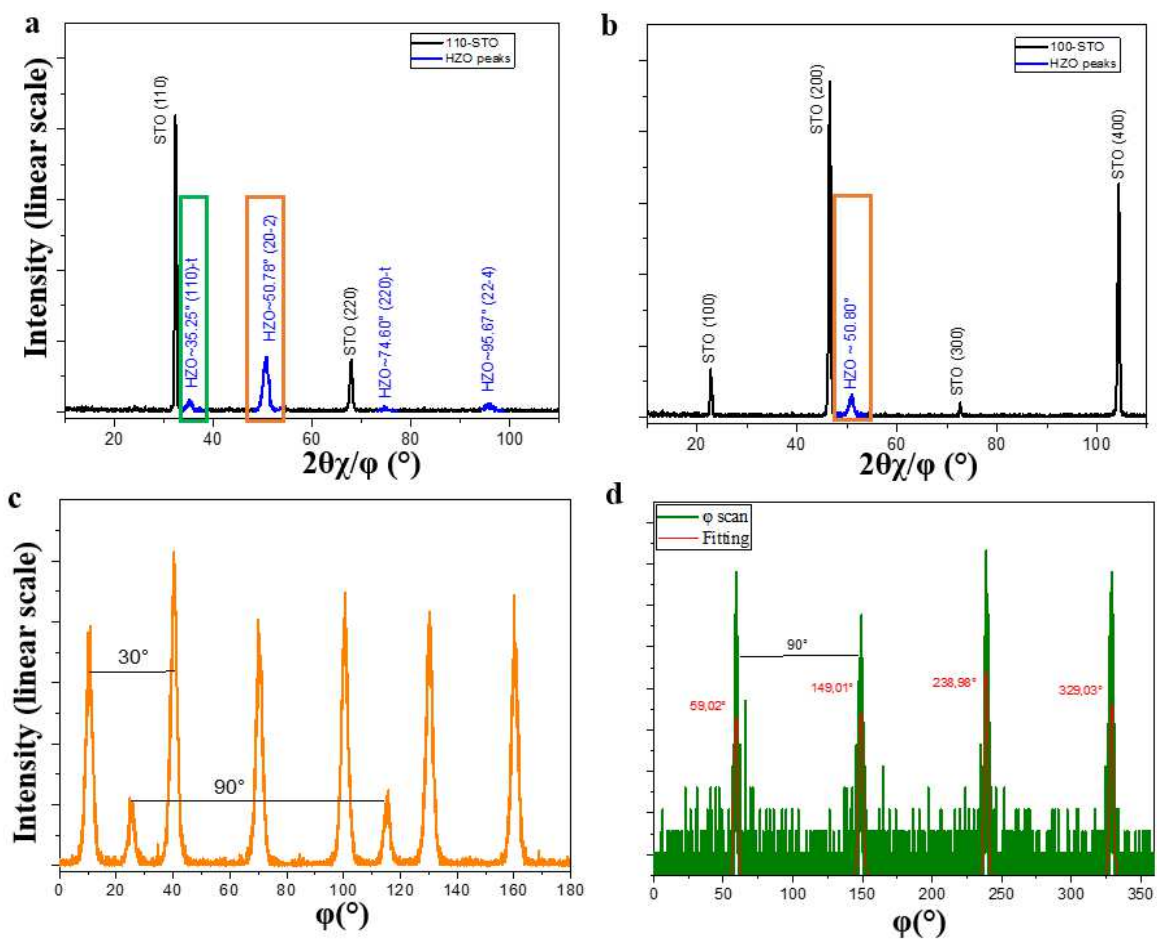

Figure S10. a, b) In-plane 2θχ/φ of 14 nm thick HZO film along <110> and <100> STO azimuthal direction respectively. c) in-plane φ -scan at 2θχ = 50.78°. d) in-plane φ -scan at 2θχ = 35.25°.

Figure S10 shows in-plane diffraction of HZO thin film on STO (001), revealing the existence of two phases, rhombohedral (Figure S10c) and tetragonal (Figure S10d). The (100) plane of the tetragonal phase is rotated by 15° with respect to the (20-2) rhombohedral. Figure S10d confirms the four-fold symmetry of the (001)-oriented tetragonal phase. Again, the t-phase is only observed in this in-plane XRD and not in out-of-plane θ-2θ (figure S8a).

**Crystallographic alignment of r-phase on 4-fold symmetry layer**

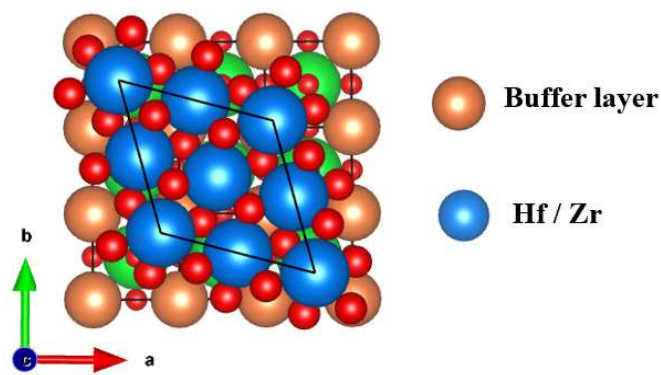

Figure S11. Possible crystallographic alignment of r-phase (3-fold symmetry) with respect to a buffer layer with 4-fold symmetry, inferred from our XRD data. The atoms at the top of the buffer and in the first layer of the rhombohedral phase are depicted in brown and blue, respectively. (a,b,c) are pseudo-cubic axes of buffer layer.

Here, we would like to present the possible alignment of the r-phase with respect to an underlining 4-fold symmetry layer, termed buffer layer for simplicity (Figure S11). This layer could be the LSMO electrode, with or without intermixed Zr/Hf atoms as reported in Ref [22], or a fully strained t-phase as observed at the interface of HZO with LSMO [9]. Note that such an ultrathin t-phase, having the same in-plane lattice parameter as the substrate, cannot be detected by in-plane diffraction scans.
Given the 4-fold symmetry of the substrate and the 3-fold symmetry of the r-phase film, the latter can have either 2 or 4 crystallographic variants [23][24]. From in-plane XRD data of Figure S7, S9, and S10, we conclude that the (20-2) rhombohedral plane ((220) in hexagonal notation) is parallel to the (110)-STO and DSO substrate ((-11-2) for DSO in orthorhombic representation). This alignment of the rhombohedral phase is depicted in Figure S11, with the long axis of the unit cell aligned along the [1-10] direction of the buffer. The same alignment is also possible in the [110] direction, giving another r-variant rotated by 90° with respect to the first one (Figure S7). For each of these two variants, another two possibilities arise from the two possible stackings in the ABC-type stacking sequence of r-phase planes in the growth direction (see Figure S2a). Finally, this gives 4 r-phase variants, as detected in the pole figure (Figure S5).